\documentclass[amsmath,amssymb,aps,10pt,prd,letterpaper,nofootinbib,balancelastpage,notitlepage,twocolumn,floatfix]{revtex4-2}
\usepackage{graphicx}	
\usepackage{ragged2e}	
\usepackage{bm}		    
\usepackage{slashed}
\usepackage[sort&compress]{natbib}	 	
	\setcitestyle{square,numbers,comma}	
\usepackage[colorlinks=true,urlcolor=blue,linkcolor=black,citecolor=red]{hyperref}	
\usepackage{xcolor}
\usepackage{fontawesome5}

\usepackage{tikz-feynman}
\usepackage{braket}
\usepackage{subfig}
\usepackage{lipsum}

\definecolor{orcidlogocol}{rgb}{0.65, 0.807, 0.223}
\newcommand{\orcid}[1]{\,\href{https://orcid.org/#1}{\textcolor{orcidlogocol}{\footnotesize\faOrcid}}}

\renewcommand{\eqref}[2][]{Eq{#1}.~(\ref{#2})}		

\usepackage{ulem}

\begin{document}

\title{Thermo-Coupled Early Dark Energy}
\date{\today}
\author{Marc Kamionkowski}
\email{kamion@jhu.edu}
\affiliation{The William H.~Miller III Department of Physics and Astronomy, Johns Hopkins University, Baltimore, MD  21218, USA}
\author{Anubhav Mathur\orcid{0000-0003-0973-1793}\,}
\email{a.mathur@nyu.edu}
\affiliation{Center for Cosmology and Particle Physics, Department of Physics, New York University, New York, NY 10003, USA}
\affiliation{School of Physics and Astronomy, Tel-Aviv University, Tel-Aviv 69978, Israel}
\begin{abstract}
Early dark energy solutions to the Hubble tension introduce an additional scalar field which is frozen at early times but becomes dynamical around matter-radiation equality. In order to alleviate the tension, the scalar's share of the total energy density must rapidly shrink from $\sim 10\%$ at the onset of matter domination to $\ll 1\%$ by recombination. This typically requires a steep potential that is imposed \textit{ad hoc} rather than emerging from a concrete particle physics model. Here, we point out an alternative possibility: a homogeneous scalar field coupled quadratically to a cosmological background of light thermal relics (such as the Standard Model neutrino) will acquire an effective potential which can reproduce the dynamics necessary to alleviate the tension. We identify the relevant parameter space for this ``thermo-coupled'' scenario and study its unique phenomenology at the background level, including the back-reaction on the neutrino mass. Follow-up numerical work is necessary to determine the constraints placed on the model by early-time measurements.
\end{abstract}

\maketitle

\section{Introduction}
\label{sec:intro}
Since the emergence of precision cosmology over the past few decades, the $\Lambda$CDM model has had considerable success at codifying the structure and evolution of the Universe in concordance with large and robust data sets, demonstrating a sophisticated understanding of systematics. Even so, a central parameter of this model, the total expansion rate of the Universe $H_0$, has a value (presently $67-68 \text{ km s}^{-1}\text{ Mpc}^{-1}$ \cite{Planck:2018vyg,eBOSS:2020yzd,DES:2021wwk}) that is discrepant with direct late-time observations (which instead suggest $ 70-75 \text{ km s}^{-1}\text{ Mpc}^{-1}$ \cite{Abdalla:2022yfr,Riess:2016jrr,Riess:2021jrx}). This tension has persisted for over a decade, leading many to consider modifications to $\Lambda$CDM that may rectify the situation. None so far is definitive \cite{Bernal:2016gxb,Verde:2019ivm,Knox:2019rjx,DiValentino:2021izs,Shah:2021onj,Efstathiou:2021ocp,Schoneberg:2021qvd}.

One promising candidate to address the $H_0$ tension is early dark energy (EDE) \cite{Kamionkowski:2022pkx,Karwal:2016vyq,Poulin:2018dzj}. This is an additional component of the Universe that behaves like a cosmological constant while the Universe is radiation dominated. Upon peaking at $\sim 10\%$ of the total energy density, it redshifts faster than radiation, becoming unobservable past the epoch of recombination. While there are many concrete realizations of this scenario, the net effect is to increase the value of the Hubble parameter that is inferred from measurements of the cosmic microwave background (CMB) and large-scale structure (LSS).

Embedding EDE into a well-defined particle physics model poses a significant challenge \cite{Gonzalez:2020fdy}. It can be emulated by a scalar field that is frozen at some initial value, becoming dynamical and thereby damping its energy density once this energy comprises a substantial fraction of the total. In order to redshift away sufficiently rapidly, the scalar must evolve in an unusually steep potential, often of exotic origin, over the brief window between matter-radiation equality (MRE) and recombination. 

In this work, we explore how such behavior may instead emerge from a simple quadratic coupling to a cosmological background of fermions, focusing on the case of Standard Model (SM) neutrinos. Such a model is one possible realization of ``thermo-coupled'' EDE (TCEDE), where the scalar potential is dominated by a thermal contribution from the fermionic sector that sharply enhances the redshifting of the net energy density as required. A secondary effect is that the vacuum expectation value of the scalar back-reacts on the neutrino mass, resulting in a distinctive phenomenology for both components as the scalar time-evolves.

These features are qualitatively similar to those found in quintessence models involving mass-varying neutrinos (MaVaNs) \cite{Fardon:2003eh,Brookfield:2005bz,Franca:2009xp,Ichiki:2007ng}, albeit with the dynamical behavior taking place around MRE rather than at late times. There is also a superficial resemblance to models such as ``neutrino-assisted'' EDE \cite{Sakstein:2019fmf,CarrilloGonzalez:2020oac,CarrilloGonzalez:2023lma}, in which scalar-neutrino interactions add a linear term to the scalar potential, and ``trigger early dark sectors'' \cite{Lin:2022phm} where the scalar instead couples to dark matter resulting in the masses becoming time-dependent. The aim in both cases is to explain why the added energy density only becomes significant at MRE (sometimes considered a coincidence problem). However, these models rely on an entirely separate bare potential to damp the scalar once it becomes active. By contrast, in this work the apparent coincidence is simply an initial condition, which may arise from misalignment or other UV physics that remains unspecified. The quadratic coupling alone suffices to reproduce the desired dynamics, and the corresponding impacts on the neutrino sector differ substantially.

The paper is organized as follows. In Sec.~\ref{sec:model}, we introduce our model, consisting of a new scalar and a fermion that is cosmologically abundant as a thermal relic. Restricting our attention to the SM neutrino, we then study the cosmological evolution of the system in Sec.~\ref{sec:evol}. We discuss the observational consequences on $H_0$ and the neutrino sector in Sec.~\ref{sec:results}, and conclude in Sec.~\ref{sec:conclusion}.

\section{Model}
\label{sec:model}

We consider the following effective Lagrangian describing a real scalar field $\phi$ that is coupled to a Dirac fermion $\psi$:
\begin{equation}\label{eq:lagrangian}
    \mathcal{L}\supset \bar{\psi}(i\slashed{\partial}-m_{\psi})\psi+\frac{1}{2}(\partial \phi)^2 -\frac{1}{2}m_{\phi}^{2}\phi^{2}-\frac{\epsilon m_{\psi}}{M_{\text{Pl}}^{2}}\phi^{2}\bar{\psi}\psi,
\end{equation}
where $\epsilon$ parametrizes the strength of the coupling and $M_{\rm{Pl}}$ is the Planck mass. The fermion $\psi$ is envisaged as a single mass eigenstate in the SM neutrino sector\footnote{This is not strictly necessary; the only requirement for thermo-coupling is that $\psi$ be cosmologically abundant up to the epoch of recombination as a thermal relic, which can be satisfied in the presence of a Majorana mass term, or even by a sterile neutrino or other dark matter candidate. However, these possibilities lead to substantial quantitative differences in the results, so for simplicity we do not consider them further in this work.}, with a bare mass $m_\psi = 0.05\text{ eV}$ consistent with existing bounds \cite{Planck:2018vyg}. The scalar is taken to be sufficiently light that it acts as a background field undergoing purely classical evolution: $m_\phi \lesssim H(z_{\rm{CMB}}) \sim 10^{-29}\text{ eV}$ where $z_{\rm{CMB}} \approx 1080$ is the redshift of recombination.

In the presence of the quadratic interaction term\footnote{Although this dimension 5 operator is suppressed at a scale $\sim M_{\rm{Pl}}^2/\epsilon m_\psi$, which is super-Planckian, it is assumed to be the leading-order interaction term. This can easily be enforced e.g. by imposing a $\mathbb{Z}_2$ symmetry on the theory.}, the fields obey the following equations of motion:
\begin{align}
   \left( i\slashed{D}- m_{\psi}- \frac{\epsilon m_\psi}{M_{\text{Pl}}^2} \phi^2 \right)\psi &= 0 \label{eq:onshell-fermion} \\
   \left(\Box + m_\phi^2 + \frac{2\epsilon m_{\psi}}{M_{\text{Pl}}^{2}}  \bar{\psi}\psi\right) \phi &= 0 \label{eq:onshell-scalar} 
\end{align}
where $D,\Box$ respectively denote the covariant derivative and d'Alembertian operators in curved spacetime. From Eq.~(\ref{eq:onshell-fermion}), we see that $\psi$ remains on-shell throughout cosmological history in the parameter regime of interest, with a scalar-dependent effective mass given by
\begin{equation}\label{eq:mpsi-eff}
    m_{\psi, {\text{eff}}} (\phi) \equiv m_\psi \left( 1 +  \frac{\epsilon \phi^2}{M_{\rm{Pl}}^2} \right).
\end{equation}
At the background level, the co-moving momentum $p_\psi/(1+z)$ is conserved due to the homogeneity of the scalar field (for an explicit computation, see Eq.~(42) of Ref.~\cite{Ichiki:2007ng}). As a result, the thermally decoupled fermions preserve a relativistic Fermi-Dirac distribution $f(p_\psi,T_\psi)$, with temperature $T_\psi \propto (1+z)$. This fixes the evolution of the fluid energy density $\rho_\psi$ and pressure $P_\psi$, allowing us to write the expectation value of $\bar{\psi}\psi$ as \cite{Smirnov:2022sfo,Bouley:2022eer}
\begin{align}
    \langle \bar{\psi}\psi\rangle &= \frac{\rho_\psi - 3P_\psi}{m_{\psi,\text{eff}}} \nonumber \\
    &= \frac{g}{(2\pi)^3} \int \frac{d^3 \mathbf{p}_\psi}{m_{\psi,\text{eff}}} \left(E_\psi - \frac{p_\psi^2}{E_\psi}\right) f(p_\psi,T_\psi) \nonumber \\
    &= \frac{m_{\psi,\text{eff}} T_\psi^2}{\pi^2} \int_0^\infty dx\,\frac{x^2 (x^2 + m_{\psi,\text{eff}}^2/T_\psi^2)^{-1/2}}{e^x + 1}, \label{eq:psibarpsi}
\end{align}
where $g=2$ is the relevant number of degrees of freedom, and $E_\psi^2 \equiv p_\psi^2 + m_{\psi,\text{eff}}^2$ is the total energy of each fermion in presence of the scalar background.

The dynamics of the scalar are captured by its effective potential, which can self-consistently be computed by combining Eqs.~(\ref{eq:onshell-scalar}) and (\ref{eq:psibarpsi}):
\begin{align}\label{eq:pot}
    \frac{1}{\phi}\frac{\partial V_{\text{eff}}}{\partial \phi} &\equiv \frac{2\epsilon m_{\psi}}{M_{\text{Pl}}^{2}}\langle \bar{\psi}\psi\rangle + m_\phi^2 \nonumber \\
    &\approx \frac{\epsilon m_\psi}{ 6M_{\text{Pl}}^{2}} T_\psi^3 \times
    \begin{cases}
     m_{\psi,\text{eff}}/ T_\psi & m_{\psi,\text{eff}} \ll T_{\psi} \\
     \sqrt{m_{\psi,\text{eff}}/T_\psi} & m_{\psi,\text{eff}} \approx T_{\psi} \\
     2.2 & m_{\psi,\text{eff}} \gg T_{\psi}
    \end{cases} \\
    &\quad + m_\phi^2. \nonumber
\end{align}
This quantity can be understood as a thermal mass for the scalar (as is familiar in the context of the QCD axion or scalar dark matter \cite{Marsh:2015xka,DiLuzio:2020wdo,Batell:2021ofv}), although it possesses a weak dependence on $\phi$ through $m_{\psi,\text{eff}}$. The thermal contribution to $V_{\text{eff}}$ will dominate as long as the bare scalar mass\footnote{While the analytic expressions include the contribution from the bare term throughout, for simplicity we set $m_\phi = 0$ in the numerical analysis.} satisfies
\begin{align}\label{eq:baremass}
    m_{\phi}^{2} &\ll \frac{\epsilon m_{\psi}^{2}T_{\psi}^{2}}{12 M_{\text{Pl}}^{2}} \\
    &\sim(10^{-29}\text{ eV})^{2}\left(\frac{\epsilon}{10^{3}}\right)\left(\frac{m_{\psi}}{0.05\text{ eV}}\right)^{2}\left(\frac{T_{\psi}}{1\text{ eV}}\right)^{2}.
\end{align}
This is comparable in magnitude to the requirement that the scalar be homogeneous up to recombination. Quantum corrections to the bare mass, arising through loops containing $\psi$, contribute at the level of \cite{Bouley:2022eer,Banerjee:2022sqg}
\begin{equation}\label{eq:finetuning}
    \delta m_\phi^2 \sim \frac{\epsilon m_\psi^2}{M_{\rm{Pl}}^2} \Lambda_{\rm{UV}}^2
\end{equation}
where $\Lambda_{\rm{UV}}$ is the cutoff scale of the theory. Naturalness of the $\phi$ mass depends on details of the UV model giving rise to the effective interactions of Eq.~(\ref{eq:lagrangian}), and other assumptions. Our model does not suffer significant fine-tuning provided that $\Lambda_{\rm{UV}}$ is at or below the eV scale.

\section{Cosmological Evolution}
\label{sec:evol}

We begin tracking the scalar field at redshift $z_0$ immediately after $\psi$  has decoupled from the SM thermal bath, which for neutrinos occurs at $z_0\approx 4\times10^9$ when their temperature is $T_0 \approx 0.7\text{ MeV}$. Since we are primarily interested in the evolution of $\phi$ around the end of radiation domination, we do not specify its behavior at earlier times, as this is model-dependent. At $z_0$, the scalar is assumed to be static and displaced from the minimum of its potential by $\phi_0$, close to the Planck scale. An initial condition of this kind is a generic expectation if $\phi$ emerges as the Goldstone boson associated with a UV symmetry that is spontaneously broken at scale $\sim \phi_0$, as in the misalignment mechanism for axion-like particles \cite{Marsh:2015xka}.

In an expanding background, the scalar evolves according to
\begin{equation}\label{eq:phieqmotion}
    \ddot{\phi} + 3 \tilde{H}\dot{\phi} + \partial_\phi V_{\text{eff}} = 0
\end{equation}
where dots denote derivatives with respect to proper time, and $\tilde{H}$ refers to the Hubble parameter, modified to include the additional energy density $\rho_{\rm{TCEDE}}$ due to the presence of this field:
\begin{equation}\label{eq:Htilde}
    \tilde{H}^2(\phi,\dot{\phi}) \equiv H^2 + \frac{8\pi }{3 M_{\rm{Pl}}^2} \rho_{\rm{TCEDE}}(\phi,\dot{\phi}).
\end{equation}
Here, $H$ is the Hubble parameter in $\Lambda$CDM, and $\rho_{\rm{TCEDE}}$ can itself be expressed in terms of field parameters by computing the canonical Hamiltonian corresponding to Eq.~(\ref{eq:lagrangian}) and subtracting the bare-mass contribution of the neutrino (which is already present in $\Lambda$CDM):
\begin{align}\label{eq:rhoEDE}
    \rho_{\rm{TCEDE}}(\phi,\dot{\phi}) &= \rho_\psi (m_{\psi,\rm{eff}}(\phi),T_\psi) - \rho_\psi (m_{\psi},T_\psi) \\
    &\qquad + \frac{1}{2}\dot{\phi}^2 + \frac{1}{2}m_\phi^2 \phi^2.  \nonumber
\end{align}
As usual, the energy density of the thermal relic $\psi$ is expressed in terms of its mass and temperature as
\begin{equation}
    \rho_\psi(m_\psi, T_\psi) = \frac{T_\psi^4}{\pi^2} \int_0^\infty dx\,\frac{x^2 (x^2 + m_\psi^2/T_\psi^2)^{1/2}}{e^x + 1}.
\end{equation}

\begin{figure}
    \centering
    \subfloat[\centering The homogeneous expectation value of the scalar field $\phi$.]{\includegraphics[width=0.95\columnwidth]{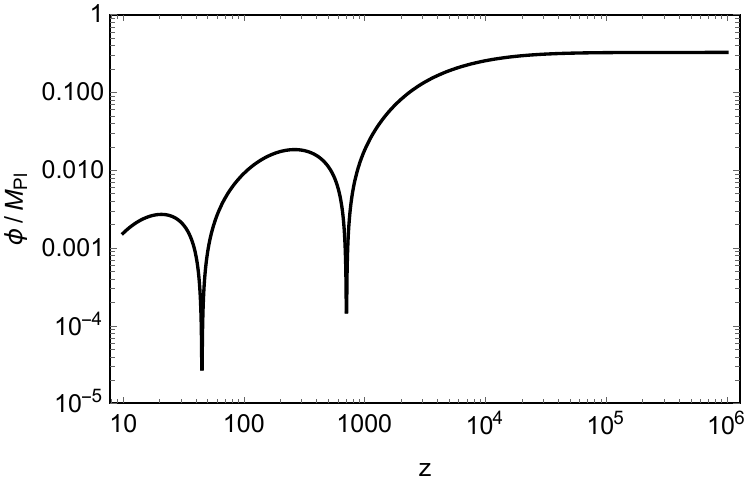}}
    \\
    \subfloat[\centering The neutrino effective mass $m_{\psi,\rm{eff}}$ in the presence of $\langle \phi \rangle \neq 0$ is shown in blue, with the bare mass $m_\psi$ and mean momentum $\langle p_\psi \rangle \approx 3.15 T_\psi$ of the sector presented for comparison in black and red, respectively.]{\includegraphics[width=0.95\columnwidth]{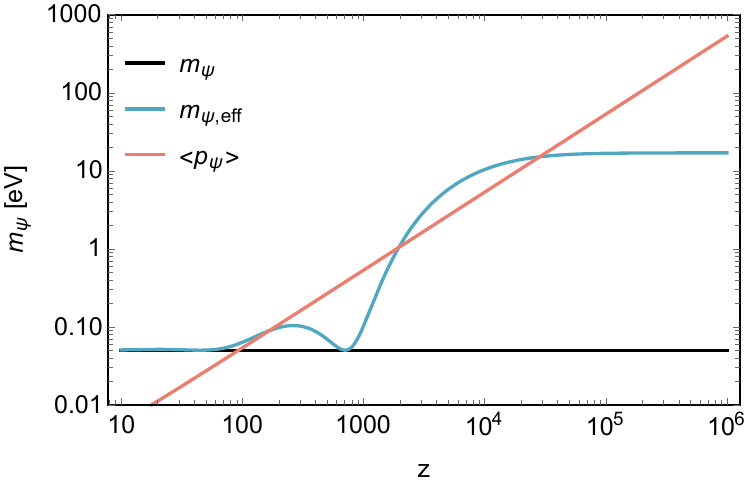}}
    \caption{Evolution of the scalar and fermion fields in the model presented in Sec.~\ref{sec:model}, for the benchmark values $\epsilon = 3 \times 10^3,\phi_0 = 4 \times 10^{27} \text{ eV}$. The bare masses are taken to be $m_\phi =  0, m_\psi = 0.05 \text{ eV}$.}
    \label{fig:field-evol}
\end{figure}

The evolution proceeds as follows. At early times, Hubble friction causes the scalar to be frozen at its initial value $\phi_0$, endowing the neutrino with a constant primordial mass $m_{\psi,\rm{eff}}(\phi_0)$. In order for $\rho_{\text{TCEDE}}$ to increase the total energy density by $f_{\text{EDE}}\sim 10\%$ near matter-radiation equality (as required to ameliorate the $H_0$ tension---see Sec.~\ref{sec:results}), we must have
\begin{equation}\label{eq:matching-edens}
    \left.\rho_{\text{TCEDE}}(\phi_0,\dot \phi = 0)\right|_{z_{\text{MRE}}} \approx f_{\text{EDE}}\frac{3M_{\text{Pl}}^2 H^2(z_{\text{MRE}})}{8\pi},
\end{equation}
which is satisfied when $m_{\psi,\rm{eff}}(\phi_0) \approx 200 m_\psi = 10\text{ eV}$, making the neutrino non-relativistic around $z_{\text{MRE}} \approx 3400$. The scalar must become dynamical starting in the same period, which (parametrically) corresponds to 
\begin{equation}\label{scalar-trigger}
    H^2(z_{\text{MRE}}) \sim \left. \frac{1}{\phi_0}\frac{\partial V_{\text{eff}}(\phi_0)}{\partial \phi} \right|_{z_{\text{MRE}}} \sim \frac{\epsilon m_\psi T_\psi^3(z_{\text{MRE}})}{M_{\rm{Pl}}^2}.
\end{equation}
Together, these conditions on the magnitude and timing of the added energy density force $\epsilon \sim 10^3$ and $\phi_0 \sim 0.1 M_{\text{Pl}}$. The precise dynamics have been computed for benchmark values and presented in Fig.~\ref{fig:field-evol}. Once $\phi$ begins to oscillate around the minimum of its potential, its decaying amplitude can be calculated in the WKB approximation as $\phi_{\rm{max}} \propto (\partial_\phi V_{\text{eff}}/\phi)^{-1/4} (1+z)^{3/2} \propto (1+z)^{3/4}$. Notice that the thermal contribution to the potential effectively slows down the damping of the oscillations compared to the standard massive scalar case. When $\epsilon (\phi_{\rm{max}}/M_{\rm{Pl}})^2 \ll 1$, close to the time that neutrinos become non-relativistic in $\Lambda$CDM, the neutrino mass returns to its bare value.

\begin{figure}
   \centering
    \subfloat[\centering The absolute energy densities, with the total energy density in $\Lambda$CDM presented in black for comparison.]{\includegraphics[width=0.95\columnwidth]{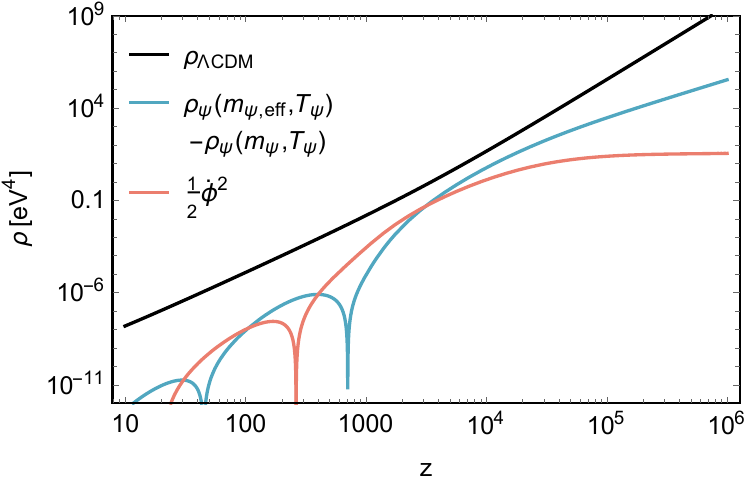}}
    \\
    \subfloat[\centering Energy densities as a proportion of the total amount in $\Lambda$CDM. The sum of the fermionic and scalar contributions is shown in purple.]{\includegraphics[width=0.95\columnwidth]{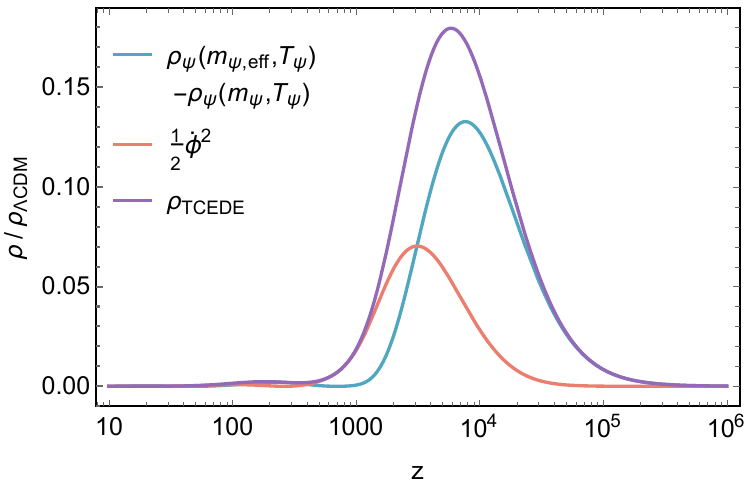}}
    \caption{Evolution of the component energy densities of the Universe in the TCEDE model of Sec.~\ref{sec:model}, for the benchmark values $\epsilon = 3 \times 10^3,\phi_0 = 4 \times 10^{27} \text{ eV}$. The bare masses are taken to be $m_\phi =  0, m_\psi = 0.05 \text{ eV}$. In both sub-figures, the kinetic energy of the scalar field is shown in red, and the additional contribution due to the modified neutrino mass is shown in blue.}
    \label{fig:rho-evol}
\end{figure}

The energy densities of the different components are shown in Fig.~\ref{fig:rho-evol}. Remarkably, at very early times when the scalar does not time evolve and the neutrino is still relativistic, $\rho_{\rm{TCEDE}}$ is \emph{not} a cosmological constant, but rather $\sim m_{\psi,\rm{eff}}^2(\phi_0) T_\psi^2 \propto (1+z)^2$. The term ``early dark energy'' is therefore something of a misnomer in the thermo-coupled scenario. Subsequently, when the scalar begins to fall toward the minimum of its potential, its kinetic energy follows $\frac{1}{2}\dot{\phi}^2 \sim  (\frac{d}{dz}\phi_{\rm{max}})^2 H^2 (1+z)^2 \propto (1+z)^{9/2}$. The same scaling is satisfied by the dominant contribution to the net neutrino energy density $ \sim m_{\psi,\rm{eff}}(\phi_{\rm{max}}) T_\psi^3 \propto (1+z)^{9/2}$. Thus, around MRE, $\rho_{\rm{TCEDE}}$ redshifts faster than any other component of the Universe including radiation, ensuring that it does not modify the post-recombination evolution from the $\Lambda$CDM baseline.

\section{Results and Discussion}
\label{sec:results}

Our model of thermo-coupled EDE has two features making it amenable to observation: the presence of additional energy density at the start of the matter-dominated era, and time-dependence in the mass of one SM neutrino species. Their combined effect is to alter the fit of the cosmological model to measurements of the cosmic microwave background (CMB), baryon acoustic oscillations (BAO), and large-scale structure (LSS). All cosmological parameters are affected simultaneously, and precisely accounting for these changes requires a numerical calculation such as with CLASS \cite{Blas:2011rf}. We leave this important analysis to future work, pointing the reader to Ref.~\cite{Ichiki:2007ng} which studies the Boltzmann equations for the general case of a neutrino interacting with a scalar at the perturbative level\footnote{We have verified that these results agree with ours at the background level. For the case of MaVaNs, similar analyses have been used to set observational constraints in e.g.~Refs.~\cite{Franca:2009xp,daFonseca:2023ury}.}\textsuperscript{,}\footnote{In addition to a dynamical effective mass, the coupling in Eq.~(\ref{eq:lagrangian}) is also responsible for long-ranged forces between neutrinos that are enhanced in the presence of a non-zero background $\phi$ \cite{VanTilburg:2024xib,Barbosa:2024tty}. These forces arise automatically from the mass variation term in the $\ell=1$ equation of Ref.~\cite{Ichiki:2007ng}, and do not need to be accounted for separately at higher multipoles, in contrast to typical analyses of neutrino self-interactions \cite{Ma:1995ey, Cyr-Racine:2013jua, Oldengott:2014qra, Taule:2022jrz}. They are expected to be weaker than gravity in the parameter space we consider.}. Instead, our focus is on identifying the viable parameter space where the Hubble tension is alleviated.

The additional energy density $\rho_{\rm{TCEDE}}$ will change the inferred value of $H_0$ through its effects on the sound horizon $r_s$ and the angular diameter distance $D_A$. The increase $\delta H_0$ can be estimated analytically as \cite{Kamionkowski:2022pkx}
\begin{align}\label{eq:delH0}
    1 + \frac{\delta H_0}{H_0} &\approx \frac{\int_{z_{\rm{CMB}}}^\infty dz\, (\rho_{\Lambda \rm{CDM}} (1 + R))^{-1/2}}{\int_{z_{\rm{CMB}}}^\infty dz\, ((\rho_{\Lambda \rm{CDM}}+\rho_{\rm{TCEDE}}) (1 + R))^{-1/2}} \nonumber \\
    &\qquad \times \frac{\int^{z_{\rm{CMB}}}_0 dz\, (\rho_{\Lambda \rm{CDM}}+\rho_{\rm{TCEDE}})^{-1/2}}{\int^{z_{\rm{CMB}}}_0 dz\, \rho_{\Lambda \rm{CDM}}^{-1/2}},
\end{align}
where $R(z) = (3/4)(\omega_{b}/\omega_{\gamma})/(1+z)$ and $\omega_b=2.24\times 10^{-2},\omega_\gamma=2.47\times 10^{-5}$ are the current physical baryon and photon densities (taken at their $\Lambda$CDM values). The factor on the second line can be neglected since $\rho_{\rm{TCEDE}} \ll \rho_{\Lambda \rm{CDM}}$ after recombination in our regime of interest. 

\begin{figure}[h]
    \centering
    \includegraphics[width=0.95\columnwidth]{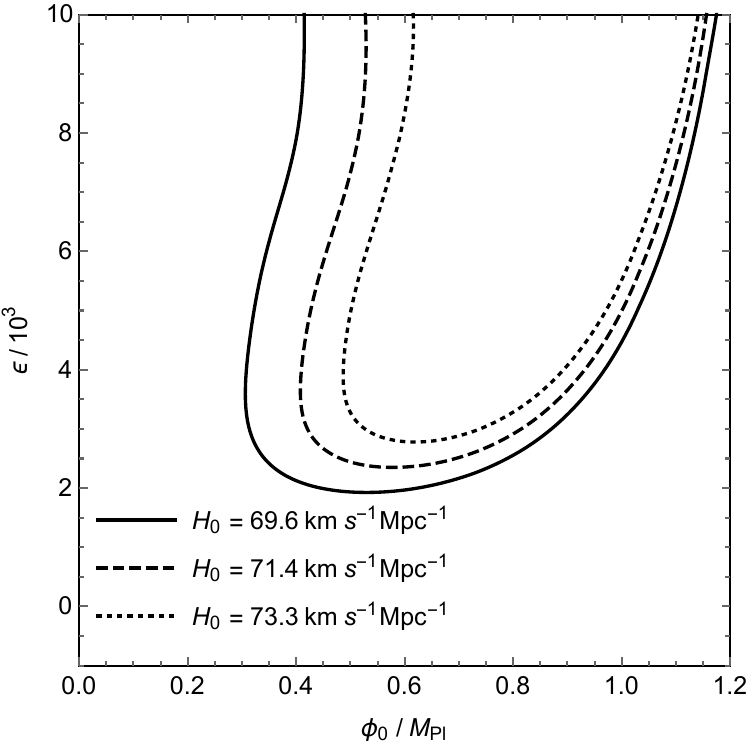}
    \caption{The parameter space for a real homogeneous scalar with an initial field value $\phi_0$ quadratically coupled to a SM neutrino with strength $\epsilon m_\psi / M_{\rm{Pl}}^2$ (see Sec.~\ref{sec:model}). Based on the estimate of Eq.~(\ref{eq:delH0}), the solid, dashed, and dotted contours correspond to reductions in the discrepancy in $H_0$ between the early-time fit of Ref.~\cite{Planck:2018vyg} and the late-time SH0ES measurement of Ref.~\cite{Riess:2021jrx}, to $4\sigma,2\sigma,0\sigma$ respectively.}
    \label{fig:par-space}
\end{figure}

Fig.~\ref{fig:par-space} maps the parameter space for the model of Sec.~\ref{sec:model}. Using $H_0 = 67.4 \text{ km s}^{-1}\text{ Mpc}^{-1}$ \cite{Planck:2018vyg} as the baseline $\Lambda$CDM value, contours demonstrate potentially viable parameter space to increase the Hubble constant to a value preferred by late-time observations. We emphasize that at those values there may be an unacceptably large effect on other cosmological parameters, or a worsening of the overall fit, so the results should be taken as indicative only. 
 
\section{Conclusion}
\label{sec:conclusion}

A scalar acting as early dark energy suggests one possible path to resolving the Hubble tension, but from a particle physics point of view, the evolution required around the time of recombination has no clear origin. In this work, we have investigated how a thermo-coupling to the SM neutrino fluid can give rise to time-dependent masses for both the scalar and the neutrino, resulting in the right scalings for the added energy density to rapidly redshift away soon after becoming relevant. We described the cosmic history of both components, and determined the values of the initial condition and interaction strength where the $H_0$ inferred from early-time data potentially aligns with measurements at late times.

A distinctive feature of our scenario is the significant periodic variation in the mass of a neutrino eigenstate throughout its evolution, which for some parameter choices even leads to a stage of non-relativistic evolution well before recombination. While the observational consequences of EDE alone are well-studied, the simultaneous inclusion of this behavior (which is an automatic consequence of the specified coupling) will leave significant imprints on the CMB and matter power spectra e.g., by hastening the onset of matter domination and modifying the free-streaming scale. This is likely to have wide-ranging impacts on the overall fit to cosmological data including at low redshifts, which may be of some import to related puzzles such as the $S_8$ tension. Future studies should therefore prioritize numerical analyses that can better delineate these effects and test the model we have presented. Should the mere two degrees of freedom present in this model prove too restrictive as-is when confronted with data, it can easily be extended by considering the presence of a Majorana mass term, or by coupling the scalar to other neutrino species, or even to a fermion in the dark sector.

\acknowledgments
We thank David E.\ Kaplan, Xuheng Luo, Ken van Tilburg, and Neal Weiner for useful discussions.  AM was supported by the Mortimer B. Zuckerman Foundation.  MK was supported by NSF Grant No.\ 2412361, NASA Grant No.\ 80NSSC24K1226, and the John Templeton Foundation.
 

\bibliography{references.bib}

\end{document}